\documentclass[twocolumn,showpacs,preprintnumbers,amsmath,amssymb]{revtex4}

\usepackage{graphicx}
\usepackage{dcolumn}
\usepackage{bm}

\begin{document}

\title{Thermally assisted magnetization reversal in the
presence of a spin-transfer torque \\
}

\author{Z. Li and S. Zhang}
\affiliation{
Department of Physics and Astronomy, University of
Missouri-Columbia, Columbia, MO 65211
}

\date{\today}

\begin{abstract}
We propose a generalized stochastic Landau-Lifshitz equation and
its corresponding Fokker-Planck equation for the magnetization
dynamics in the presence of spin transfer torques. Since the spin
transfer torque can pump a magnetic energy into the magnetic
system, the equilibrium temperature of the magnetic system
is ill-defined. We introduce an effective temperature based on a
stationary solution of the Fokker-Planck equation. In the limit of
high energy barriers,
the law of thermal agitation is derived. We find that the
N\'{e}el-Brown relaxation formula remains valid as long as we
replace the temperature by an effective one that is linearly dependent of
the spin torque. We carry out the numerical integration of the stochastic
Landau-Lifshitz equation to support our theory. Our results agree
with existing experimental data.

\end{abstract}

\pacs{75.60.Jk, 75.75.+a}

\maketitle

\section{Introduction}

Thermally assisted magnetization reversal has been the subject of
intensive theoretical and experimental study for many decades.
Aside from the relevance of the subject to emerging magnetic
technology such as heat-assisted magnetic recording \cite{heat}
and thermal stability of magnetic random access memory
\cite{MRAM}, fundamental physics of magnetization reversal process
driven by white-noise is very rich. Classical transition-rate
theory of Kramer \cite{Kramer} has supplied a framework in
understanding thermal activation of a single domain magnetic
element \cite{Brown}. Namely, the thermal switching probability
$P(t)$ can be described by the N\'{e}el-Brown (NB) relaxation-time
formula, $P(t) = 1 - \exp(-t/\tau)$, where the relaxation time is
$\tau = f_0^{-1} \exp( E_b/k_B T)$, $f_0$ is an attempt frequency,
$E_b$ is the energy barrier, and $T$ is the temperature. For a
multi-domain structure, the energy surface becomes extremely
complicated and identifying energy barriers are numerically
non-trivial. Nevertheless, with recent development of
micromagnetic modeling, one can understand thermal reversal
reasonably well for a not-too-complicated structure \cite{Koch}.

An implicit and yet critical assumption in the NB theory is that
magnetization dynamics is governed by a torque from
an effective magnetic field ${\bf H}_{eff} = - \mbox{\bf $\nabla$}_{\bf M}
E({\bf M})$, where $E({\bf M})$ is the total magnetic energy, i.e., the
effective field is derivable from the derivative of an energy function
with respect to the magnetization vector. Therefore, an energy barrier
is well-defined in the NB relaxation formula. If the torque is not derivable
from an energy function, one would expect breakdown of the NB formula.
Recently, a new class of torques, called spin transfer torque (STT),
has been proposed \cite{Berger,Slon} and verified experimentally
\cite{Katine,Kent}. STT is derived from a spin polarized current in magnetic
multilayers. For a spin valve structure, STT is written as \cite{Slon}
\begin{equation}
\mbox{\boldmath $\Gamma$}_s=\frac{\gamma a_J}{M_s}
{\bf M} \times ({\bf M} \times
\hat{\bf M}_p)
\end{equation}
where $a_J$ represents the strength of STT; it is proportional to
the current density. $\gamma$ is gyromagnetic ratio, $\hat{\bf
M}_p$ is a {\em unit} vector representing the direction of the
magnetization of the pinned layer, ${\bf M}$ is the magnetization
vector of the free layer and $M_s = |{\bf M}|$ is the saturation
magnetization. If we define an effective field ${\bf H}'_J \equiv
(a_J/M_s) {\bf M} \times \hat{\bf M}_p $ from STT, it is evidently
that ${\bf H}'_J$ can {\em not} be written as a total derivative
of a function with respect to the magnetization vector, i.e.,
there is no well-defined energy associated with the field ${\bf
H}'_J$.

Recent experiments \cite{Buhrman,Bass1,JEW1} on the thermal
effect of the spin torque had also indicated that the thermally
assisted magnetization reversal can not be simply described by the
NB formula. Urazhdin {\em et al} \cite{Bass1} found that the
activation energy strongly depends on the magnitude as well as the
direction of the current. To capture the gloss features of
the observed experiments, they have to introduce an effective temperature
unrelated to the true temperature in the NB formula.
Their proposed effective temperature was
then interpreted via a possible magnetic heating and magnetic
excitations from the spin transfer torque. The
current directional dependence of the effective temperature
indicated that the heating is not of the ordinary current-induced
Joule heating. However, no attempt has been made to mathematically
link the effective temperature with the spin transfer torque of
Eq.~(1).

The problem of thermally assisted escape process driven by a
non-gradient driven force, not derivable from a potential, is an
unresolved outstanding problem in statistical physics. While there
are already some efforts to formulate the escape time in this case, the general
conclusion is that the law of escape time lacks universality and a
variety of scaling relations exist \cite{Stein}. The standard
treatment of the thermal escape problem in the presence of a
non-gradient field would start from the Fokker-Planck (FP)
equation and one numerically solves for the probability
distribution \cite{Stein}. This procedure involves proper
averaging over the possible escape trajectories. Let us consider
the total work done by the conservative torque and the
non-conservative STT along an arbitrary trajectory,
\begin{eqnarray}
\delta W  = - \int ({\bf H}_{eff} + {\bf H}'_{J}) \cdot d{\bf M}
\nonumber\\
 = E_b - \frac{a_J}{M_s} \int_{{\bf M}_0}^{{\bf M}_f} (
{\bf M} \times {\bf e}_x ) \cdot d{\bf M}
\end{eqnarray}
where we have assumed that the magnetization vector starts at an
initial equilibrium point ${\bf M}_0 = M_s {\bf e}_x $ and reaches
to an energy saddle point ${\bf M}_f$, and we have defined the
energy barrier from the conservative torque $E_b = E({\bf M}_f) -
E({\bf M}_0)$. One immediately realizes that $\delta W$ defined
above depends on the escape trajectory. In the absence of STT, one
relies on the assumption that the most probable path of the escape
is through a minimum energy barrier, i.e., ${\bf M}_f $ would be
an energy saddle point. In the presence of STT, such assumption
breaks down in general. In the present paper we do not intend to
address the general problem of the thermal escape in an open
system, rather we want to answer a very focused question: to what
extent one can formulate the thermal agitation in terms of a
simplified N\'{e}el-Brown activation process and to what accuracy
one can analyze the relevant experimental data through a simple
effective formula we will develop in the later sections. The paper
is organized as follows. In Sec.~II, we propose the stochastic
Landau-Lifshitz equation and its corresponding Fokker-Planck
equation in the presence of the current. In Sec.~III, we introduce
a stationary solution for the probability density of magnetization
by identifying an effective barrier or an effective temperature
associated with spin torques. In Sec.~IV, we present a numerical
calculation to demonstrate the validity of our theory in several
realistic cases. Finally, we compare our theory with existing
experiments and summarize our theory.

\section{Stochastic Landau-Lifshitz equation in the presence of currents}

Let us first explicitly propose the following generalized
stochastic Landau-Lifshitz (LL) equation that describes dynamics
of the magnetization vector subject to a STT at finite
temperatures
\begin{equation}
\label{LLG} \frac{d{\bf M}}{dt}= - \gamma {\bf M} \times ({\bf
H}_{eff}+{\bf h}_r)- \frac{\gamma\alpha}{M_s} {\bf M}\times [{\bf
M} \times ({\bf H}_{eff} + {\bf h}_r)] + \mbox{\boldmath $\Gamma
$}_s,
\end{equation}
where $\alpha$ is the damping constant, ${\bf H}_{eff}$ is the
effective magnetic field including the external field, the
anisotropy field, the exchange field, and the demagnetization
field, and ${\bf h}_r$ is a fluctuating field with a Gaussian
stochastic process whose statistical properties are defined as
\begin{equation}
<h_r^i (t)> =0; \hspace{0.25in} <h_r^i (t)h_r^j (t')> =
2D\delta_{ij} \delta (t-t')
\end{equation}
where $i$ and $j$ are Cartesian indices, $D$ represents the
strength of the thermal fluctuations whose value will be
determined later. $< >$ denotes an average taken over all
realization of the fluctuating field. In the absence of the spin
torque, the above equation is the standard stochastic LL equation.
Note that we have conveniently dropped the customary renormalized
gyromagnetic ratio $\gamma/(1+\alpha^2)$ when compared with the
standard Landau-Lifshitz-Gilbert equation. The critical assumption
of our proposed stochastic LL equation, Eq.~(3), is that {\em the
spin torque does not contain a fluctuating field ${\bf h}_r$}. The
justification for this choice is that the spin torque comes from
the conduction electrons whose transport properties are less
affected by thermal fluctuations since the Fermi level is much
higher than the thermal energy. Therefore, the thermal fluctuation
would not appear to affect $a_J$ which represents the strength of
the spin torque. We believe that our proposed
stochastic LL equation captures the main random processes induced
by thermal fluctuation. Nevertheless, one could, in principle,
have introduced a
second random field or torques to take into account the
fluctuation of the spin torque. In our proposed LL equation, the
thermal effect on the spin torque is only encoded in the
dependence of the magnetization vector.

To establish the thermal properties from the above stochastic
equation, one must take a proper thermal average. Fortunately,
much of theoretical work on the stochastic LL in the absence of
the spin torque had been carried out \cite{Brown,Lazaro}. Here we
will follow and generalize the procedure pioneered by Brown
\cite{Brown}. We define $P({\bf M},t)$ as a non-equilibrium
probability density for magnetic orientation vectors associated
with the stochastic Eq.~(3). The rate equation for $P({\bf M},t)$
is
\begin{equation}
\frac{\partial P}{\partial t} + \mbox{\boldmath $\nabla$} \cdot
{\bf J} -\lambda \nabla^2 P =0
\end{equation}
where the probability current density ${\bf J}= P d{\bf M}/dt$,
$\mbox{\boldmath $\nabla$}$ is a short notation for the gradient
operator on the magnetization vector $\mbox{\boldmath
$\nabla$}_{\bf M}$, and $\lambda$ is the diffusion constant whose
value is determined by fluctuation-dissipation theorem. In the
present case, $\lambda$ is related to $D$ by $\lambda=(1/2)D
\gamma_0 (1+\alpha^2) $. The above rate equation, Eq.~(5), is a
simple statement for the angular momentum conservation: the change
of probability density in an enclosed small volume (first term)
has to be balanced by the net probability in-flowing flux (second
term) plus the probability density loss via spin diffusion (third
term). By inserting Eq.~(3) into Eq.~(5), after a straightforward
but rather tedious algebra manipulation \cite{footnotes}, the
resulting equation is the Fokker-Planck equation
\begin{equation}
\begin{array}{lll}
\frac{\partial P}{\partial t} =
 - \mbox{\boldmath $\nabla$}
\cdot \left\{ \left[-\gamma {\bf M }\times {\bf H}_{eff} + {\bf
\Gamma}_s -\frac{\gamma \alpha}{M_s} {\bf M }\times ({\bf M
}\times {\bf H}_{eff}) \right . \right. \\
  \left . \left . + \gamma^2 (1+\alpha^2)D
{\bf M }\times ({\bf M }\times \mbox{\boldmath $\nabla$} ) \right]
P \right\}
\end{array}
\end{equation}
In the absence of the spin torque, i.e., ${\bf \Gamma}_s =0$, the
thermal equilibrium distribution density $P$ demands to take the
form of the Boltzmann distribution function, i.e., $P(a_J=0, T)
\propto \exp(-E/k_{B}T)$ where $T$ is the temperature and $E$ is
the energy defined by $H_{eff} = - \mbox{ \boldmath $\nabla$} E$.
Inserting this equilibrium $P(a_J=0, T)$ into Eq.~(6), one finds
that
\begin{equation}
D= \frac{\alpha}{1+\alpha^2} \cdot \frac{k_BT}{\gamma M_s};
\end{equation}
this is the well-known dissipation-fluctuation relation.
We now postulate that {\em the fluctuating field is independent of the
spin torque}. This hypothesis is consistent with our notion that the
spin torque is a deterministic action so that the spin torque does not
alter the randomness induced by the thermal fluctuation. With this identification,
the stochastic LL equation, Eq.~(3), and its corresponding Fokker-Planck
equation, Eq.~(6),
completely determine the dynamics of the magnetization vector at finite
temperature $T$.

\section{Stationary solution and effective temperatures}

Before we numerically solve the above stochastic LL equation for a
number of interesting cases, we should first look for a stationary
solution in Eq.~(6), i.e., $P=P_0 ({\bf M}) $ is independent of
time. Without the spin torque, this solution is known as the
equilibrium Boltzmann distribution function mentioned earlier.
With the spin torque, the system is no more in an equilibrium
state because the system is subject to the spin torque and thus it
is not a closed system. For an open system, the law of thermal
dynamics does not require the minimum free energy and the concept
of thermal equilibrium breaks down. Nevertheless, it is still
meaningful to obtain a stationary solution $P_s$ where
Fokker-Planck probability density is time-independent $\partial
P_s / \partial t = 0$. Thus,
\begin{equation}
\begin{array}{lll}
\mbox{\boldmath $\nabla$} \cdot \left\{ \left[-\gamma {\bf M
}\times {\bf H}_{eff} + \Gamma_s -\frac{\gamma \alpha}{M_s} {\bf M
}\times ({\bf M }\times {\bf H}_{eff}) \right.\right.\\
\left.\left. + \gamma^2 (1+\alpha^2)D {\bf M }\times ({\bf M
}\times \mbox{\boldmath $\nabla$} ) \right] P_s \right\} =0
\end{array}
\end{equation}
Unfortunately, the above eigenstate problem for an arbitrary field
${\bf H}_{eff}$ is generally difficult to solve. One can
immediately verify that the Boltzmann probability density $P_0
\propto \exp(-E/k_BT)$ is no more a solution of the above
equation. To make progress, we need to consider a special case
below. First, we recall that the N\'{e}el-Brown formula for the
thermal agitation is in fact most useful where the energy barrier
constructed by ${\bf H}_{eff}$ is much higher than the thermal
energy $k_BT$. In this limit, the probability density will be very
small if the direction of the magnetization vector is away from
the energy minimum. Here we should also consider this case. We now
tentatively seek a solution of $P_s$ in the form of $P_s \propto
\exp(-E/k_BT^*)$ where we have introduced an effective temperature
$T^*$ that will be determined next. By placing this $P_s$ into Eq.~(8)
and by utilizing the identity $\mbox{\boldmath $\nabla$} \cdot ({\bf M }\times
{\bf H}_{eff}P_s ) \propto \mbox{\boldmath $\nabla$} \cdot ({\bf M
}\times \mbox{\boldmath $\nabla$}P_s )=0 $, we find
\begin{equation}
\begin{array}{lll}
\mbox{\boldmath $\nabla$} \cdot \left\{ \left[ \alpha \left(
\frac{T}{T^*} - 1 \right) {\bf M}\times ({\bf M}\times {\bf
H}_{eff}) \right. \right.\\
\left.\left. + a_J {\bf M}\times ({\bf M}\times {\bf M}_p )
\right] P_s \right\} =0
\end{array}
\end{equation}
Clearly, the above equation does not necessarily have a solution for
an arbitrary effective field. However, as we point out earlier,
we have limited ourselves to a high barrier
case such that the magnetization vector at the stationary condition is
nearly at the direction of ${\bf M}_p$. For concreteness, let us consider
a most experimentally relevant geometry where
${\bf M}_p = {\bf e}_x$ and
\begin{equation}
{\bf H}_{eff} = (H_{ext} + \frac{H_K}{M_s} M_x) {\bf e}_x - 4 \pi
M_z {\bf e}_z
\end{equation}
where $H_{ext}$ is the external field which is applied at
x-direction, $H_K$ is the anisotropy field, and $- 4 \pi M_z {\bf
e}_z$ is the demagnetization field perpendicular to the plane of
the film. In this case, the energy minimum are at $M_x = \pm M_s$,
$M_y=M_z =0$. We simply keep the first order in $M_y$ and $M_z$,
and set $M_x = \sqrt{M_s^2 -M_y^2-M_z^2} = M_s$ up to the first
order, we find
\begin{eqnarray}
{\bf M}\times ({\bf M}\times {\bf H}_{eff}) \approx M_y (H_{ext} +
H_K) {\bf e}_y \nonumber\\
 + M_z (H_{ext}+H_K + 4\pi M_s) {\bf e}_z
\end{eqnarray}
and $ {\bf M}\times ({\bf M}\times {\bf M}_p ) \approx
M_y {\bf e}_y + M_z {\bf e}_z$. By placing these expressions into Eq.~(9) and
carrying out the divergence $\mbox{\boldmath $\nabla$}$ for
$M_y$ and $M_z$ components we find (one noticed that the divergence operator
on $P_s$ produces higher orders since Eq.~(11) is already
the first order in $M_y$ and $M_z$, thus it is consistent with our
approximation by neglecting the terms of divergence on $P_s$),
\begin{equation}
\alpha (2 \pi M_s + H_{ext}+H_K) \left( \frac{T}{T^*} -1 \right) +
a_J = 0
\end{equation}
or
\begin{equation}
T^* = T \left( 1-\frac{a_J}{a_c} \right) ^{-1}
\end{equation}
where we defined a critical spin torque $a_c = \alpha (H_{ext}+H_K
+ 2 \pi M_s)$. Coincidentally, this critical spin torque is
precisely the minimum spin torque required to switch the
magnetization at zero temperature \cite{Sun}.

We should point out that the concept of the effective temperature
introduced here should be understood in terms of the stationary
solution of the probability density. The thermally averaged
dynamical variable, for example, the magnetization vector $<{\bf
M}> = \int P_s {\bf M} \sin\theta d\theta d\phi$ would behave as
if the temperature of the system is $T^*$. However, the magnetic
temperature which is defined through the thermal fluctuation
remains to be $T$. An alternative understanding of this effective
temperature is to rewrite the stationary solution by $P_0 \propto
\exp(-E^*/k_BT)$ where $E^*$ is an effective activation energy
\begin{equation}
E^* = E \left( 1- \frac{a_J}{a_c} \right) .
\end{equation}
In other word, we can state that the spin torque alters the
magnetic energy and thus there will be an effective energy barrier
associated with the spin current. Therefore, it is equivalent to
think of the effect of the spin torque via the modification of the
temperature or of the energy barrier.

To conclude this section, we have found a stationary solution
$P_s$ of the stochastic LL equation. Since the life time or the
relaxation time $\tau$ is inversely proportional to the
probability density $P_s$, we can write the generalized
Brown-N\'{e}el formula below
\begin{equation}
\tau^{-1} = f_0 \exp \left(
- \frac{E_b (
1- {a_J}/{a_c} )
}{k_BT}
\right)
\end{equation}
where $f_0$ is an attempt frequency and $E_b$ is the energy barrier.

\section{Comparison between numerical and analytical Results}

The stationary solution, $P_s \propto \exp[-E({\bf M})/k_BT^*]$,
is based on the assumption of high
energy barrier assumption. In general cases, one should start the
calculation of magnetization dynamics from our generalized LL equation,
Eq.~(3). Since the stationary solution is simple and easy to use in
analyzing experimental data, it would be necessary to establish the range
of validity of Eq.~(15) for various interesting experimental situations.
Once its validity is established, we expect that
our stationary solution would be serving as a starting point to understand
various thermal agitation phenomena in the present of the current.
In this Section, we numerically integrate Eq.~(3) and compare the result
with $P_s$.

\begin{figure}
\centering
\includegraphics[width=7cm]{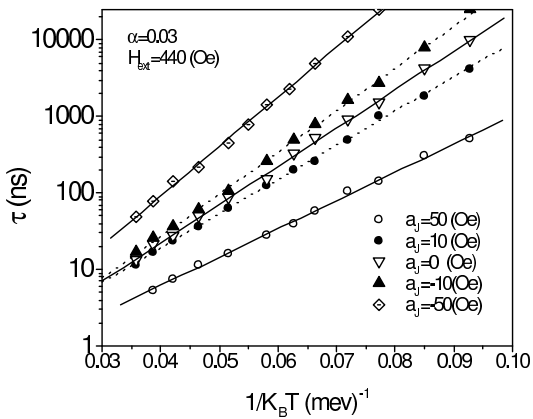}
\caption{Relaxation time $\tau$ as a function of the inverse of
the temperature for several values of the spin-transfer torques.
The external field of $H_{ext}=440$ (Oe) is applied along $-x$
direction.}
\end{figure}

The standard white-noise spectrum, Eq.~(4) where $D$ is given by
Eq~(7), is used for the modeling of the temperature dependence of
random fields. The calculation procedure is same as that for the
standard LLG equation, except a spin torque is added to the
equation of the motion. A magnetic layer, whose lateral size is
64nm$\times$64nm and whose thickness is 2.5nm, is treated as a
single macrospin. The in-plane uniaxial anisotropy field $H_K$ is
$500$(Oe) and the saturation magnetization is $4 \pi M_s = 12,000$
(Oe). These parameters are reasonably consistent with the
experiments performed by the Cornell group \cite{Katine}. The
Gilbert damping constant was taken as $\alpha=0.03$ throughout the
modeling. The magnetization of the free layer is initially
saturated at $+x$ direction. At $t=0$, we apply a magnetic field
at $-x$ direction whose magnitude is close to but less than the
anisotropy field $H_{K}$. At the same time, $a_{J}$ is also
applied to the system. Equation (3) was numerically integrated in
time using the stochastic Heun method with a 0.3ps time step. A
smaller time step has been tested and it yields nearly identical
results in all the cases presented in the paper.

With above specified parameters and procedure, we first determine
the probability $P(t)$ that the magnetic layer has been reversed
within the waiting time $t$. By performing up to $5\times 10^4$
independent runs for each set of parameters (each run starts at
$t=0$ and ends at the time that the magnetization has either been
just switched or ends at the time up to $t = 5 \mu s$, whichever
is smaller) and then by recording the number of them that the
magnetization is switched at time interval ($t$, $t+\Delta t$), we
obtained a simulated $P(t)$ that is fitted by a simple exponential
function, i.e., $P(t) = 1- \exp(-t/\tau )$ where $\tau$ is the
fitting parameter for the relaxation time. We have found that the
fit works remarkably well for any values of $a_J$ we have
considered. In Fig.~1, we show the fitted relaxation time $\tau$
as a function of the temperature for a fixed applied field
$H_{ext}=440 Oe$. Two features are immediately seen: first the
data fall on a straight line for any fixed $a_J$; this indicates
the thermally assisted reversal can be described by an activation
process, i.e., $ \ln \tau \propto 1/k_B T$. Therefore, it is
meaningful to introduce an effective activation energy, see
Eq.~(14), in accordance with the N\'{e}el-Brown law of thermal
agitation. The second conclusion is that the slope, or the
activation energy depends on STT: the positive $a_J$ favors a
lower energy barrier. All these features are well described by
Eqs.~(13)-(15).
\begin{figure}
\centering
\includegraphics[width=7cm]{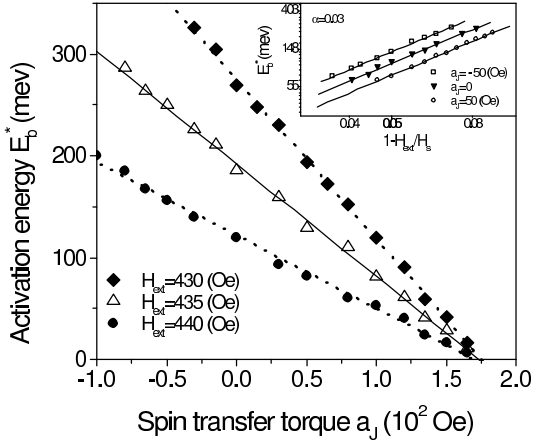}
\caption{Activation energy $E_b^*$ as a function of $a_{J}$ at
three different external fields. Inset: $E_b^*$ vs $1-H_{ext}/H_s$
for several different values of STT.}
\end{figure}
In Fig.~2, the effective activation energies are shown to be
linearly dependent on the current and they vanish at almost the
same point $a_c$ for different external fields [note that $a_c$ is
weakly dependent on the external field, see the definition of
$a_c$ after Eq.~(13)]. In the insert of Fig.~2, we have shown the
activation energy as a function of the magnetic field for several
different STT. The activation energy can be fitted by

\begin{eqnarray}
E_b^* =  E_{b} (H_{ext}) \left( 1-\frac{a_J}{a_{c}}
\right)\nonumber\\
 =E_{0} \left( 1-\frac{H_{ext}}{H_s}
\right)^{\beta} \left( 1- \frac{a_J}{a_{c}} \right)
\end{eqnarray}
where $H_{s}$ is the switching field at zero temperature, $E_{0}$
is the energy barrier at zero magnetic field field, and $\beta$ is
a constant, which has been argued to be 1.5 or 2. The exponent
$\beta = 2$ for the external field applied parallel to the easy
axis. These simulated results confirm our analytical result,
Eq.~(14).

\section{Comparison with experimental data}

A number of experiments on spin torque induced thermal agitation
had been carried out. It would be interesting to see whether our
prediction, Eq.~(15), agrees with these existing data. The
phenomenon that we want to compare first is so-called ``telegraph noises''
or dwell times. Experimentally, one simultaneously applies an
external magnetic field and a spin current to a spin valve
structure so that the magnetization direction of the free layer is
fluctuating from one direction to another due to thermal agitation
\cite{Bass1,JEW2}. The dwell time $\tau_P$ $(\tau_{AP})$ is
defined as an average time the magnetization of the free layer is
parallel (antiparallel) to that of the fixed layer. In general,
$\tau_P \neq \tau_{AP}$. However, by adjusting the magnetic field
or the spin current, one is able to obtain an equal dwell time for
parallel and antiparallel states, $\tau_P = \tau_{AP}$. From
Eq.~(15) for the parallel and the antiparallel states, the condition of the
equal dwell time is
\begin{equation}
\left( 1+ \frac{H_{ext}}{H_s} \right) ^{1.5} \left(
1-\frac{I}{I_{c}^{AP}} \right) = \left( 1- \frac{H_{ext}}{H_s}
\right) ^{1.5} \left( 1-\frac{I}{I_{c}^P} \right)
\end{equation}
where $I_{c}^{AP}$ and $I_{c}^P$ are the critical currents for the
magnetization switching from antiparallel to parallel alignments
and vice versa; their magnitudes are not necessary the same, i.e.,
$I_{c}^{AP} \neq -{I_{c}^P}$ in a typical experimental geometry
\cite{Buhrman,JEW1}. To compare our prediction, Eq.~(17), with
experiments, we plot the H-I phase diagram of equal dwell time
$\tau_P = \tau_{AP}$ in Fig.~3.

\begin{figure}
\centering
\includegraphics[width=7cm]{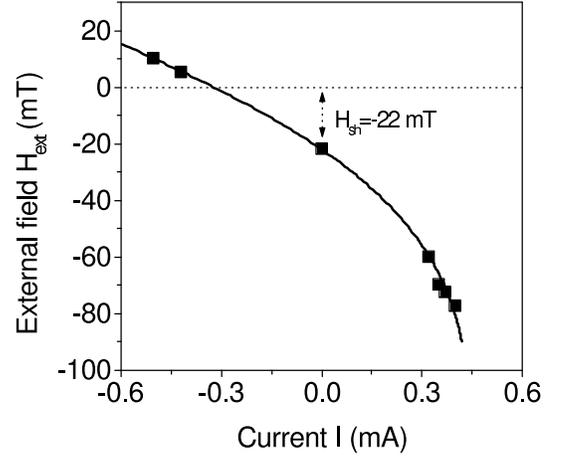}
\caption{H-I Phase boundary of equal dwell times
$<\tau_{AP}>=<\tau_{P}>$. The coupling field is $H_{sh}=-22 mT$,
the critical currents are $I_{c}^{AP}=-1.25 mA$ for the transition
from AP to P alignments and $I_{c}^P=0.425 mA$ from P to AP
alignments. Line: Eq.~(17) except $H_{ext}$ being replaced by
$H_{ext}+H_{sh}$. Square: experimental data \cite{JEW2}.}
\end{figure}

It is noted that we have shifted the external field by $H_{sh}=
-22 (mT)$ to take into account of the magnetic coupling between
the free and fixed layers. The coupling may come from either the
exchange or dipolar couplings. Since we assume that the
magnetization of the fixed layer is held at the direction of ${\bf
e}_x$, the free layer receives an effective coupling field that
will be added to $H_{ext}$ in Eq.~(17). Evidently, the agreement
between our results and the experimental data is excellent
\cite{JEW1,JEW2}.

Next we compare the ratio of the dwell times of antiparallel and parallel
states for a fixed magnetic field. Again, from Eq.~(15), we have
\begin{equation}
\frac{\ln (f_{0} \tau_P)}{\ln (f_{0} \tau_{AP})} = \frac{ \left(
1- \frac{H_{ext}}{H_s} \right) ^{1.5} \left( 1-\frac{I}{I_{c}^{P}}
\right) }{ \left( 1+ \frac{H_{ext}}{H_s} \right) ^{1.5} \left(
1-\frac{I}{I_{c}^{AP}} \right) }
\end{equation}

\begin{figure}
\centering
\includegraphics[width=7cm]{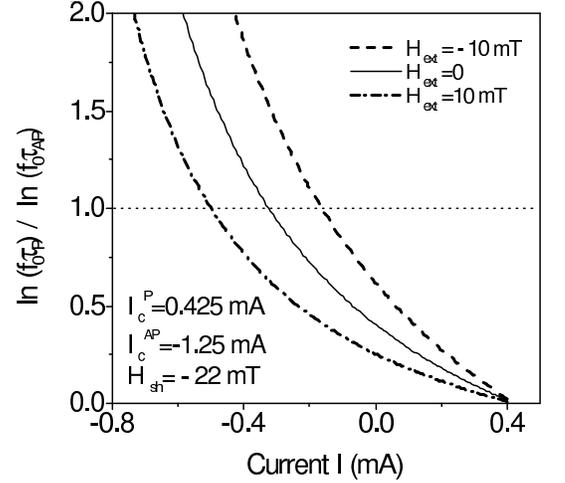}
\caption{Ratio of the relaxation times $\ln (f_0\tau_{P})/\ln (f_0
\tau_{AP})$. The parameters are same as those in Fig.4.}
\end{figure}

In Fig.~4, we show the ratio of the dwell times for a fixed
magnetic field as a function of the spin torque by using the same
set of experimental parameters as in Fig.~3. The results are
consistent with experimental data (however, the data points in
Ref.\cite{JEW1,JEW2} are rather scattered so that I do not include
those data in the figure).

Up till now, we have studied the thermal activation by {\em abruptly}
introducing an external field and a STT at t=0. In experiments,
there are ramping times, e.g., STT is gradually increasing
at rate of, say $10^{-5} Oe/\mu s$ \cite{Buhrman} and
care must be taken when one
compares our theory, Eq.~(15), with experiments. In the current ramping
period, $a_J$ is not a constant and thus the activation energy,
Eq.~(16), varies with time. In this case, one should utilize the differential
form of the switching probability instead of $P(t)=1-\exp (-t/\tau)$,
\begin{equation}
\frac{dP(t)}{1-P(t)} = \frac{dt}{\tau(t)}.
\end{equation}
By assuming a linear ramping of STT, i.e., $C = \frac{da_J}{dt}$
is a constant and by placing Eqs.~(15) and (16) into Eq.~(19), we integrate
Eq.~(19) from $t=0$ to $t=t_0$ and find the average switching STT
$a_c (T,C) \equiv a_J (t_0)$
\begin{equation}
a_c (T, C) \cong a_c \left[
1-\frac{k_{B}T}{E_{b}} \ln \left(
\frac{f'_{0}k_{B}Ta_c}
{E_{b} C } \right)
\right]
\end{equation}
where $f'_0 = - f_0 \ln[1-P(t_0)]$.
The variance of the switching STT is found as
\begin{equation}
\sigma_a \cong a_c \frac{k_{B} T}{E_{b}}
\end{equation}
These relations, Eqs.~(20) and (21),
are consistent with earlier studies on the similar
energy barrier formalism for a very different physical system
\cite{Kurkijarvi}.

\begin{figure}
\centering
\includegraphics[width=7cm]{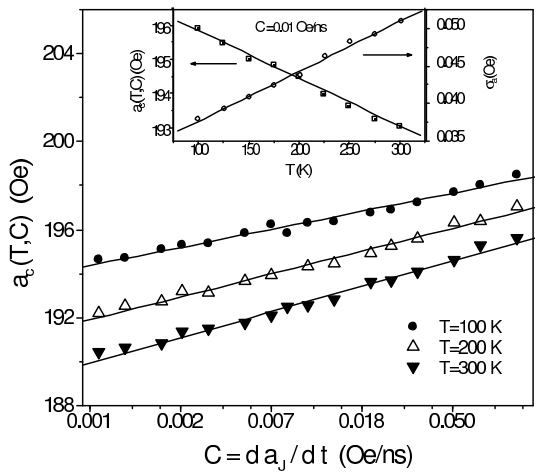}
\caption{Dependence of $a_c (T,C)$ on the sweeping rate $C$ at
finite temperature. Inset: $a_c (T,C)$ and $\sigma_a$ as a
function of temperature for the sweeping rate of STT at $0.01
Oe/ns$.}
\end{figure}

Numerically, the finite ramping rate can be rather easily handled.
In determining $a_c (T, C)$ at finite temperature, we ramp $a_J$
with a fixed rate. At a certain time, the magnetization vector
switches and we record the value of $a_J$. By repeating the above
procedure 800 times, we are able to establish the switching $a_J$
histograms from which the mean switching $a_c (T,C)$ and its
standard deviation $\sigma_a$ are obtained.  At temperature
between $100 K$ and $300 K$, we calculated the mean and standard
deviation of the distributions at a sweeping rate between $0.001
Oe/ns$ and $0.1 Oe/ns$. As expected for a thermally activated
process, $a_c (T,C)$ increases with decreasing temperature and
with increasing sweeping rate. Figure ~5 shows temperature and
sweeping rate dependence of $a_c (T,C)$ and $\sigma_a$.

A logarithmic dependence of $a_c (T,C)$ on the sweeping rate has
been found, which is in a good agreement to Eq.(20). Moreover, the
inset of Fig.~5 describes the temperature dependence of $a_c
(T,C)$ and $\sigma_a$. We have verified that the temperature
dependence of $\sigma_a$ is a linear relationship and $a_c (T,C)$
monotonically decreases with increasing $T$.

Myers {\em et al} \cite{Buhrman} have discovered that the thermal
activation driven by spin torque is qualitatively different from
that driven by the magnetic field. They have suggested an
activation energy whose form is similar to ours, except that they
have postulated an arbitrary exponent, i.e., $\delta W \propto
(1-a_J/a_c)^{\xi}$. Although they have stated that ${\xi}$ might
be 1.5, most of experimental data shown in the paper can be used
to determine the value of $\xi $. One set of data, Fig.~1(d) of
Ref.~[11], shows that $a_c (T, C) $ linearly increases with $\ln
C$ as predicted by our Eq.~(15). If one uses different scaling
relation, e.g., $\delta W \propto (1-a_J/a_c)^{1.5}$, one would
obtain $a_c (T, C) \propto (\ln C)^{2/3}$ that would disagree with
the experimental data. Therefore, the existing data support the
linear scaling between the activation energy and the spin torque.

\section{Conclusions}

In summary, we have extended the law of thermal agitation to include
the spin transfer
torque driven by the spin polarized current in magnetic multilayers.
Although the concept of the energy barrier or the temperature
in the N\'{e}el-Brown formula
breaks down in the presence of the spin transfer torque, we are able to
re-establish the N\'{e}el-Brown formula by properly introducing an
activation energy or an effective temperature
to replace the true energy barrier or true lattice temperature.
Our formalism is further supported by numerical solutions and is
in agreement with experimental results.

This work is supported by NSF (ECS-0223568).


\begin{thebibliography}{1}

\bibitem{heat}
J. J. M. Ruigrok, R. Coehoon, S. R. Cumpson, and H. W. Kestern,
J. Appl. Phys. {\bf 87}, 5398 (2000).

\bibitem{MRAM}
J. M. Daughton, J.  Appl. Phys. {\bf 81}, 3758 (1997).

\bibitem{Kramer}
H. A. Kramer, Physica {\bf 7}, 284 (1940).

\bibitem{Brown}
W. F. Brown, Phys. Rev. {\bf 130}, 1677 (1963).

\bibitem{Koch}
R. H. Koch, G. Grinstein, G. A. Keefe, Y. Lu, P. L. Trouilloud,
W. J. Gallagher, and S. S. P. Parkin, Phys. Rev. Lett.
{\bf 84}, 5419 (2000).

\bibitem{Berger}
L. Berger, Phys. Rev. B {\bf 54}, 9353 (1996); {\bf 59}, 11465 (1999).

\bibitem{Slon}
J. Slonczewski, J. Magn. Magn.
Mater. {\bf 159}, L1 (1996); {\bf 195}, L261(1999).

\bibitem{Katine}
J. A. Katine, F. J. Albert, R. A. Buhrman,
E. B. Myers, and D. C. Ralph, Phys. Rev. Lett. {\bf 84}, 3149 (2000).

\bibitem{Kent}
B. \"{O}zyilmaz, A. D. Kent, D. Monsma, J. Z. Sun, M. J. Rooks,
and R. H. Koch, Phys. Rev. Lett. {\bf 91}, 067203 (2003).

\bibitem{Buhrman}
E. B. Myers, F. J. Albert, J. C. Sankey, E. Bonet, R. A. Buhrman,
and D. C. Ralph, Phys. Rev. Lett. {\bf 89}, 196801 (2002)

\bibitem{Bass1}
S. Urazhdin, Norman O. Birge, W. P. Pratt, Jr., and J. Bass, Phys.
Rev. Lett. {\bf 91}, 146803 (2003); S. Urazhdin {\it et al.} Appl.
Phys. Lett. {\bf 7}, 114 (2003)

\bibitem{JEW1}
J. E. Wegrowe, Phy. Rev. B {\bf 68}, 2144xx (2003).

\bibitem{Stein}
R. S. Maier and D. L. Stein, Phys. Rev. Lett. {\bf 69}, 3691
(1992).

\bibitem{Lazaro}
J. L. Garc\'{i}a-Palacios and F. J. L\'{a}zaro, Phys. Rev. B {\bf
58}, 14937 (1998).

\bibitem{footnotes}
we can rewrite the stochastic LL Eq.~(3) in a general form of a
system of Langevin equations with multiplication noise.

\bibitem{Sun}
J. Z. Sun, Phys. Rev. B {\bf 62}, 570 (2000); Ya. B. Bazaliy, B.
A. Jones, and S. Zhang, J. Appl. Phys. {\bf 89}, 6793 (2001); Z.
Li and S. Zhang, Phys. Rev. B {\bf 68}, 024404 (2003).

\bibitem{JEW2}
A. Fabian, C. Terrier, S. Serrano Guisan, X. Hoffer,
M. Dubey, L. Gravier, J.-Ph. Ansermet, J.-E. Wegrowe,
cond-mat/0304191.

\bibitem{Kurkijarvi}
J. Kurkij\"{a}rvi, Phys. Rev. B {\bf 6}, 832 (1972).
\end{thebibliography}
\end{document}